\documentclass[
 aip,
 jmp,
 amsmath,amssymb,
 reprint,
]{revtex4-1}

\usepackage{graphicx}
\usepackage{dcolumn}
\usepackage{bm}

\begin{document}

\preprint{AIP/123-QED}

\title
{Electromagnetic Waves in a Model with Chern-Simons
Potential}

\author{D.Yu. Pis'mak}
 \altaffiliation[Also at]{ Institut for Theoretical Physics, ETH-Zurich.}
\author{Yu.M. Pis'mak}
\altaffiliation[Also at]{ Institut for Theoretical Physics,
Ruprecht Karls University Heidelberg.}
\affiliation{ Department of Theoretical Physics, Saint Petersburg
State University
}

\author{F. J. Wegner}
\affiliation{
Institut for Theoretical Physics, Ruprecht Karls University Heidelberg
}

\date{\today}

\begin{abstract}

We investigate the appearance of Chern-Simons terms in electrodynamics
at the surface/interface of materials.
The requirement of locality, gauge invariance
and renormalizability in this model is imposed.
Scattering and reflection of electromagnetic waves in three different homogeneous
layers of media is determined.
Snell's law is preserved. However, the transmission and reflection
coefficient depend on the strength of the Chern-Simons interaction,
and parallel and perpendicular components are mixed.

\end{abstract}

\pacs{12.20.Ds, 11.15.Yc, 41.20.Jb, 42.25.Gy, 68.60.Wm}

\keywords{Casimir Effect, Chern-Simons potential, QED in singular background,
Nanophysics}

\maketitle

\section{Introduction }

Space-time homogeneity  and isotropy are typical for usual quantum
field theory models of elementary particles. It is a natural
assumption in the study of various processes with simplest
excitations of quantum vacuum.  However, it is not suitable for
modelling the interaction of quantum fields with macroscopic
objects, changing essentially  the vacuum properties. In this
case, quantum macro-effects may appear in dynamics of material
bodies which can not be explained in the framework of  classical
physics. Theoretically, this problem was first considered in 1948
by Casimir, who showed that quantum vacuum fluctuations cause the
attraction between two perfectly conducting parallel plates of an
uncharged capacitor \cite {First}. This phenomenon, called the
Casimir effect (CE), is observed experimentally, and the results
obtained empirically for perfectly conductive materials are with a
high degree of accuracy in agreement with theoretical ones \cite
{Second, Third, Forth, Five2}. At typical distances of 10-1000 nm for the
CE both  quantum and classical features of the system  become
essential. Their combination forms a special nano-physics.
Investigations of it are not only of general theoretical interest.
 They are important also for the development of new technical devices,
 in view of the increasing trend towards their miniaturization.

Although there are numerous papers devoted
 to the theoretical problems of the CE \cite {Five2, Five},  they are based often
  on simplified models of a free scalar field theory with fixed boundary
conditions, applying  only to investigations of some particular aspect of the CE, and
ignoring  usually specificity of quantum electrodynamics.
Such models are not suitable
for a complete
description of a wide range of nano-physical phenomena occurring in the system
as a result of the
interaction of quantum degrees of freedom with the material body of a given
shape (classic defect).
The results presented in our paper were obtained within the Symanzik
approach \cite{Five1}
for construction
 of quantum field theory models when there are spatial inhomogeneities with
sharp boundaries.
They are described by an additional action functional (action of the defect)
that is
concentrated in the region of space where the macroscopic object is located.
In quantum
  electrodynamics the interaction of photons with the  defect modelling
background field
   is completely determined by the requirements of the locality, gauge
invariance, renormalizability,
    and is described by the Chern-Simons action functional with a dimensionless
constant characterizing
    the material properties of the surface \cite {Six}.
It affects the Casimir
force, which is non-universal
and can be not only attractive, but also repulsive for a flat capacitor \cite {Six}.
It is shown also that in this model
     the static electric charge interacting with the surface defect generates a
magnetic field, and stable straight-line
     current creates  an electric field \cite {Six}.
The calculated Casimir-Polder potential for a neutral atom near a flat surface
allowed to find the parity-violating
 corrections to the previously known results \cite{Seven}.
 Based on  the earlier proposed model \cite{Six} we study in
 this paper    the
electromagnetic waves  in three layers of  matter with magnetic
susceptibilities $\mu_1,\ \mu_2, \ \mu_3$ and permittivities
$\varepsilon_1,\ \varepsilon_2,\ \varepsilon_3$ separated by two parallel
material planes $x_3=\pm l/2$ whose Chern-Simons interaction with the
electromagnetic field is
characterized by coupling constants $a_1, \ a_2$.

\newcommand{\ie}{{\rm i}}
\newcommand{\ex}[1]{{\rm e}^{#1}}
\newcommand{\de}{{\rm d}}
\newcommand{\mat}[1]{\left(\begin{array}{cc} #1 \end{array}\right)}
\newcommand{\mbfA}{{\bf K}}
\newcommand{\mbfD}{{\bf L}}

\section{Statement of problem }

For the formulation and investigation of the model
it is convenient to use the notation $\check{\alpha}$, and
$\textbf{a}$ for three- and  two-component arrays correspondingly.
We define also the scalar product and  $\ast$-composition of them:
$\check{\alpha}
\check{\beta}=\alpha_1\beta_1+\alpha_2\beta_2+\alpha_3\beta_3$,
$\textbf{a}\textbf{b}=a_1b_1+ a_2b_2$, $\check{\alpha}\ast
\check{\beta}=(\alpha_1\beta_1,\alpha_2\beta_2,\alpha_3\beta_3)$,
$\textbf{a}\ast\textbf{b}=(a_1b_1, a_2b_2)$.

Let us introduce the arrays
\begin{eqnarray*}
 \check{\theta}_l\equiv (\theta(-l/2-x_3),\theta(l/2-|x_3|),\theta(x_3-l/2)),\
\\
\textbf{d}_l\equiv(\delta(x_3+l/2),\delta(x_3-l/2)).
\end{eqnarray*}
Here $\theta(\alpha)$ and $\delta(\alpha)$ are Heaviside step-function and Dirac
delta-function.
The scalar products of $\check{\theta}_l$ with
$\check{\beta}=(\beta_1,\beta_2,\beta_3)$
and $\textbf{d}_l$ with $\textbf{c}=(c_1,c_2)$ are defined as
\begin{eqnarray*}
{\cal F}(\beta_1,\beta_2,\beta_3)={\cal F}(\check{\beta})\equiv
\check{\beta}\check{\theta}_l, \\
  {\cal D}(c_1,c_2)={\cal D}(\textbf{c})\equiv \textbf{c} \textbf{d}_l.
\end{eqnarray*}
Then one obtains
\begin{eqnarray}\label{F1}
    \frac{\partial}{\partial x_3}{\cal F}(\check{\beta})
= {\cal F}\left(\frac{\partial}{\partial x_3}\check{\beta}\right)
+{\cal D}(\textbf{s}({\check{\beta}})),\ \ \nonumber \\\label{F2}
     {\cal F}(\check{\beta}){\cal F}(\check{\gamma}) ={\cal
F}(\check{\beta}\ast\check{\gamma}), \
{\cal F}(1,1,1)=1. \ \ \ \nonumber
\end{eqnarray}
 where $\textbf{s}(\check{\beta})\equiv (\beta_2-\beta_1,\beta_3-\beta_2)$.
The  model \cite{Six} of the photon field $A_\mu$ interacting with the
two-dimensional material surface described by equation $\Phi(x)=0$
can be generalized for the considered system by the definition of the
action functional as
\begin{equation}\label{0}
    S(A) =-\frac{1}{4}G_{\mu\nu}F^{\mu\nu}+S_{\phi}(A).
\end{equation}
Here, $F_{\mu\nu}\equiv\partial_\mu A_\nu - \partial_\nu A_\mu$,
$G_{\mu\nu}\equiv {\cal E}(x_3)F_{\mu\nu}$, if $\mu=0$ or $\nu=0$,
and $G_{\mu\nu}\equiv {\cal M}^{-1}(x_3)F_{\mu\nu}$ if $\mu\neq 0$, $\nu\neq 0$
with
$ {\cal E}(x_3)\equiv {\cal F}(\check{\varepsilon})$,
${\cal M}(x_3)\equiv {\cal F}(\check{\mu})$.

 The functional $S_{\phi}(A)$ describes the interaction of
the 2-dimensional material objects (defects) with the photon field.
The defects lie in our case at two parallel planes $x_3=l_i$
with ${\bf l}=(-l/2,+l/2)$.
Using the notation $\Phi_j(x)=x_3-l_j$ we can write the action of
the defects as $S_\phi(A)=S_1(A)+S_2(A)$ where
\begin{eqnarray}\label{sd}
  S_j(A)=
    \frac{a_j}{2}\int \partial_{\mu}\Phi_j(x)A_{\nu}(x)
    \tilde{F}^{\mu\nu}(x)\delta(\Phi_j(x))\de x= \nonumber \\
    = \frac{a_j}{2}\int A_{\nu}(x)
    \tilde{F}^{3\nu}(x)\delta(\Phi_j(x))\de x, \ j=1,2.
\nonumber
\end{eqnarray}
In (\ref{sd}) $\tilde{F}^{\mu\nu}$ is the dual field
tensor
$\tilde{F}^{\mu\nu}=\epsilon^{\mu\nu\lambda\rho}F_{\lambda\rho}$,
and $\epsilon^{\lambda\mu\nu\rho}$ is the totally antisymmetric
tensor, $\epsilon^{0123}=1$.

The Euler-Lagrange equations for the action functional $S(A)$ (\ref{0})
are written as modified Maxwell's equations,
\begin{equation}\label{EL}
  \frac{\delta S(A)}{\delta A_{\nu}} = \partial_{\xi}G^{\xi \nu}+ {\cal
D}(\textbf{a})  J^\nu=0.
\end{equation}
We use the notations $ J^\nu \equiv
 \epsilon^{3 \nu \sigma \rho}F_{\sigma \rho}$, $\textbf{a}\equiv (a_1,a_2)$.
We construct the general solution of eqs. (\ref{EL}),
analyze its properties and consider processes  of plain wave scattering.

Action (\ref{0}) and the Euler-Lagrange equations (\ref{EL})
 are invariant under gauge transformation
 $A_{\mu}(x)\rightarrow A_{\mu}(x)+\partial_\mu
 \varphi(x)$. Thus the solution of (\ref{EL}) is
 defined up to a gauge transformation. We fix it by choosing the temporal
gauge $A_0= 0$. Then the vector-potential $A^\mu =(0,\vec{A})$
yields the electric field $\vec{E}=-\partial_0\vec{A}$ and the
magnetic induction $\vec{B}=\vec{\partial}\times \vec{A}$.

We solve the eqs. (\ref{EL}) using the Fourier transform over
coordinates $x_0=ct, x_1, x_2$ for the vector-potential $A_\mu$:
\begin{eqnarray}
\label{four1}
   A_{\mu}(x) = \frac{1}{(2\pi)^{\frac{3}{2}}} \int \ex{\ie
   \overline{p}\overline{x}}A_{\mu}(x_{3}, \overline{p})
\de\overline{p}= \nonumber \\
   = \frac{2 \Re}{(2\pi)^{\frac{3}{2}}} \int \theta(p_0)
\left[\ex{\ie \overline{p}\overline{x}}A_{\mu}(x_{3},
   \overline{p})\right]\de\overline{p}\nonumber
\end{eqnarray}
Here and later we use the notation $\overline{p}$ for vector
$\overline{p}=(p_0,p_1,p_2)$,
$\overline{p}\overline{x}=p_0x_0-p_1x_1-p_2x_2$.
$\Re$ denotes the real part and $\omega =cp_0$ the frequency.

\section{Solution of Euler-Lagrange equations}

With the gauge condition $A_0=0$, the eqs. (\ref{EL}) for
 $\vec{A}(x_{3}, \overline{p})$ are equivalent to the following ones
\begin{eqnarray}\label{eqrho}
 (\partial_{3}{\cal E}{\cal P}^{-2}\partial_{3} + {\cal E})\rho =
   \frac{2\, \ie}{p^0}{\cal D}(\textbf{a})\tau,
\\ \label{eqtau}
(\partial_{3}{\cal M}^{-1}\partial_{3}+{\cal M}^{-1}{\cal
P}^2)\tau =-2\,\ie p^{0}{\cal D}(\textbf{a})\rho, \\ \label{eqa3}
 A_3 = -{\cal P}^{-2}\partial_{3}\rho
\end{eqnarray}
where
\begin{eqnarray}
  \rho\equiv\ie p_1A_1+\ie p_2A_2,\
\tau\equiv \ie p_2 A_1-\ie p_1 A_2, \nonumber \\
  {\cal
P} \equiv{\cal F}(\kappa_1,\kappa_2,\kappa_3),\ \kappa_i\equiv
\sqrt{p_0^2\varepsilon_i\mu_i -p_1^2-p_2^2}. \nonumber
\end{eqnarray}
By definition the real part of $\kappa_j$ is chosen to be non-negative, and if
it vanishes, then $\kappa_j=-\ie|\kappa_j|$.

The fields $\rho,\ \tau $ are found from eqs.
(\ref{eqrho}, \ref{eqtau}). The components
$A_1$, $A_2$ of the vector-potential $\vec{A}$ are expressed by
$\rho$ and $\tau$,
\begin{eqnarray}
\label{VPot} A_1= -\ie ( \rho\, p_1  + \tau\, p_2 )\mathfrak p^{-2}, \
A_2=\ie ( \tau\, p_1-\rho\, p_2)\mathfrak p^{-2},
\end{eqnarray}
where $\mathfrak p^2 =p_1^2+p_2^2$.
The electromagnetic field $\vec{A}(x_3,\bar{p})$ in the considered medium is
characterized by the
mutually orthogonal vectors $\vec{p}_\|=(p_1,p_2,0)$,
$\vec{p}_{\bot}=(p_2,-p_1,0)$,
$\vec{t}=(0,0,1)$.
The vectors $\vec{p}_\|$, $\vec{t}$ define the plane of incidence.
In virtue of (\ref{eqa3}), (\ref{VPot}), the vector potential
$\vec{A}=(A_1,A_2,A_3)$
can be presented in the form $\vec{A}=\vec{A}_\|+\vec{A}_\bot$ where
$\vec{A}_\|$ is parallel
to the plane of incidence,  and $\vec{A}_\bot$ is perpendicular to it,
\begin{eqnarray}\label{pres}
\vec{A}_\|(x_3,\bar{p})= \left(-\ie \vec{p}_\|\mathfrak{p}^{-2}
 -\vec{t}\, {\cal P}^{-2}\partial_3\right)\rho(x_3,\bar{p}),\\ \label{pret}
 \vec{A}_\bot(x_3,\bar{p})= -\ie\vec{p}_\bot\mathfrak{p}^{-2}\tau(x_3,\bar{p}).
\end{eqnarray}
Since in our gauge $\vec{E}(\bar{p},x_3)=-\ie p_0
\vec{A}(\bar{p},x_3)$, the field $ \rho(x_3,\bar{p})$, ($
\tau(x_3,\bar{p})$) describe plane waves whose electric field
vectors are parallel (perpendicular) to the plane of incidence.
Eqs. (\ref{eqrho}, \ref{eqtau}) show that the Chern-Simons defects
mix parallel and transverse components of the phonon field.

Let us introduce the notations ${\bf f}(x_3)=(\rho(x_3),\tau(x_3))$ and define
\begin{eqnarray}
\nonumber
\mbfA=\left(\begin{array}{cc}
  {\cal E}{\cal P}^{-2} & 0 \\
  0 & {\cal M}^{-1} \end{array}\right),\
 \mathbf{C}=\left(\begin{array}{cc}
  0 & p_0^{-1}\\
  -p_0 & 0 \end{array} \right), \\
\nonumber
\mbfD_i=\left(\begin{array}{cc}
  e_i & 0\\
  0 & m_i \end{array}\right),\
e_i=\frac{\varepsilon_i}{\kappa_i},\ m_i=\frac{\kappa_i}{\mu_i},\ i=1,2,3.
\end{eqnarray}
Then we can present (\ref{eqrho}, \ref{eqtau}) in a
compact form
\begin{eqnarray}\label{eqPhi}
 (\partial_{3}\mbfA\partial_{3} + \mbfA{\cal P}^2){\bf f} =
   2 \ie{\cal D}(\textbf{a}) {\bf C}{\bf f}.
\end{eqnarray}
We conclude that ${\bf f}$ is continuous at $x_3=l_j$,
\begin{equation}\label{cond1}
{\bf f}_j(l_j) = {\bf f}_{j+1}(l_j),
\end{equation}
since a discontinuity would yield
a $\delta'$-function on the l.h.s. of (\ref{eqPhi}), which is absent on
the r.h.s. Due to (\ref{VPot}) $A_{1,2}$ is continuous at
the defects. Thus the derivatives $\partial_{0,1,2}A_{1,2}$ are
continuous, which implies the continuity of the components
$E_{1,2}$ and $B_3$.

Introducing ${\bf f}(x_3)={\cal F}(\check{\bf f}(x_3))$ with
$\check{\bf f}(x_3)=({\bf f}_1(x_3), {\bf f}_2(x_3), {\bf f}_3(x_3))$
we integrate (\ref{eqPhi}) from $x_3=l_j-\eta$ to
$x_3=l_j+\eta$ with infinitesimal $\eta$
\begin{equation}\label{eqPhiint}
\frac{\mbfD_{j+1}}{\kappa_{j+1}}\partial_3 {\bf f}_{j+1}(l_j)
-\frac{\mbfD_j}{\kappa_j}\partial_3 {\bf f}_j(l_j) = 2\ie a_j{\bf
C}{\bf f}(l_j).
\end{equation}

Within the layers $x_3\not=\pm l_j$ eq. (\ref{eqPhi}) is
written as $(\partial^2_3+\kappa_i^2){\bf f}_i(x)=0$ and yields
\begin{equation}\label{solution}
{\bf f}_i={\bf f}_i^+ + {\bf f}_i^-,\ {\bf f}_i^\pm=
(\rho^\pm_i,\tau^\pm_i) = {\bf c}_i^\pm e^{\mp i\kappa_i x_3}.
\end{equation}
For real $\kappa_i$ the solution with the upper (lower) sign
describes a plane wave moving in positive (negative)
$x_3$-direction.

It follows from (\ref{solution}), (\ref{pres}), (\ref{pret})
that $\vec{A}=\vec{A}^++\vec{A}^-$ and
\begin{eqnarray}\label{amppar}
\vec{A}_\|^\pm(x_3)= -\frac{\ie
\vec{p}_\|^{\,\pm}\rho^\pm(x_3)}{\mathfrak{p}^{2}},\
\vec{A}_\bot^\pm(x_3)= -\frac{\ie \vec{p}_\bot
\tau^\pm(x_3)}{\mathfrak{p}^{2}}
\end{eqnarray}
with
\begin{eqnarray}\label{amport}
 \vec{p}_\|^{\,\pm} \equiv \vec{p}_\| \mp \mathfrak{p}^2 {\cal P}^{-1}
\vec{t},\
 (\vec{p}_\|^{\, \pm})^2={\cal P}_0^2{\cal P}^{-2}\mathfrak{p}^2,  \\
\label{amport1}
 {\cal P}_0\equiv {\cal P}|_{p_1=p_2=0}=p_0{\cal F}(n_1,n_2,n_3), \
n_i=\sqrt{\varepsilon_i\mu_i}.
\end{eqnarray}

Since $\partial_3 \textbf{f}_j=\ie\kappa_j \tilde{\textbf{f}}_j$,
where $\tilde{\textbf{f}}_j\equiv \textbf{f}^-_j-\textbf{f}^+_j $,
the conditions (\ref{eqPhiint}) can be written as
\begin{eqnarray} \label{cond1a}
\mbfD_{j+1}\tilde{\textbf{f}}_{j+1}(l_j) -
\mbfD_j\tilde{\textbf{f}}_j(l_j) =2a_j
\mathbf{C}\textbf{f}_j(l_j), \ j=1,2.
\end{eqnarray}
These eqs. describe the discontinuity of the components
$H_{1,2}$ of the magnetic field and $D_3$ of the dielectric displacement
due to the currents $a_j J^{\nu}$ in (\ref{EL}),
\begin{eqnarray} \label{discD}
D_{3,j+1} - D_{3,j} = -a_j J^0_j=-2a_jB_{3,j}, \\
H_{1,j+1} - H_{1,j} = -a_j J^2_j=2a_jE_{1,j}, \\ H_{2,j+1} -
H_{2,j} = a_j J^1_j=2a_jE_{2,j}. \label{discH}
\end{eqnarray}

In order to solve the eqs.  (\ref{cond1}, \ref{cond1a}) it
is convenient to introduce the following $2\times 2$ matrices
\begin{eqnarray}
\mathbf{T}^{\alpha \beta}_j=
\mathbf{1}+\alpha\mbfD_{j+1}^{-1}(\beta\mbfD_j-2a_j\mathbf{C}), \
j=1,2, \ \alpha,\beta=\pm 1 \nonumber
\end{eqnarray}
and 4-component vectors $\mathbf{U}_j=
(\mathbf{u}^+_j,\mathbf{u}^-_j)$, $\mathbf{V}_j=
(\mathbf{v}^+_j,\mathbf{v }^-_j)$ with
$\mathbf{u}^{\pm}_j=\mathbf{f}_j^{\pm}(l_j)$,
$\mathbf{v}^{\pm}_j=\mathbf{f}_{j+1}^{\pm}(l_j)$. Then we obtain
from (\ref{cond1},\ref{solution},\ref{cond1a}) the relations
between the $\bf V$ and $\bf U$ by means of the transfer matrices
$T$,
\begin{eqnarray} \nonumber
  \mathbf{V}_j=T_j\mathbf{U}_j,\  \mathbf{U}_2=T_l \mathbf{V}_1,\
\mathbf{V}_2=T\mathbf{U}_1,\ T= T_2T_lT_1,  \\ \nonumber
 T_l=\left( \begin{array}{cc}
   \ex{-\ie l\kappa_2}\mathbf{1}& 0 \\
  0 & \ex{\ie l\kappa_2}\mathbf{1}
\end{array} \right),\
 T_j=\frac{1}{2}\left( \begin{array}{cc}
  \mathbf{T}^{++}_j & \mathbf{T}^{+-}_j \\
  \mathbf{T}^{-+}_j  & \mathbf{T}^{--}_j
\end{array} \right).
\end{eqnarray}

One has for nonactive media (real $\varepsilon$, $\mu$ and $a$)
\begin{eqnarray}\label{MatRel}
G_j= T_j^{\dag} G_{j+1} T_j,\ T_l^{\dag}G_jT_l=G_j, \ T^\dag
G_3T=G_1, \nonumber \\ \label{Pointing} \mathbf{U}_1^*
G_1\mathbf{U}_1=\mathbf{V}_1^{*}G_2\mathbf{V}_1=
\mathbf{U}_2^{*}G_2\mathbf{U}_2=\mathbf{V}_2^{*}G_3\mathbf{V}_2.
\
\end{eqnarray}
Here $\dag$, $*$ denote the hermitian conjugation of matrix and
the complex conjugation of  vector components,
\begin{eqnarray}
\nonumber G_j \equiv \mat{ \frac{\Re \kappa_j }{|\kappa_j|}{\bf
g}_j & \frac{\Im \kappa_j} {|\kappa_j|}{\bf g}_j \\ -\frac{\Im
\kappa_j}{|\kappa_j|}{\bf g}_j
 & -\frac{\Re \kappa_j }{|\kappa_j|}{\bf g}_j},\ {\bf g}_j
 \equiv \mat{ p_0 e_j & 0 \\ 0 & m_j/p_0
 },
\end{eqnarray}
$\Re \kappa _j\ (\Im \kappa_j )$ is the real (imaginary) part of
$\kappa_j$.

For a complete analysis of the propagation of waves in the considered
medium it is enough to assume that in the region $x_3>l/2$ there
are no waves moving in the negative direction of $x_3$-axis. This
restriction obeys $\mathbf{f}_3^-(l/2)=\mathbf{v}_2^-=0$, since
for real $\kappa_3$, $\mathbf{f}_3^-(l/2)$ is the amplitude of the
wave moving from $x_3=+\infty$ to the plane $x_3=l/2$ , and for
imaginary $\kappa_3$ the field must decay exponentially for
$x_3\rightarrow+\infty$. Then $\mathbf{V}_2=T\mathbf{U}_1$ yields
\begin{eqnarray}\label{eqamp}
 \mathbf{T}^{-+}\mathbf{u}_1^+ +\mathbf{T}^{--}\mathbf{u}_1^-=0, \
\mathbf{v}^+_2=\mathbf{T}^{++}\mathbf{u}_1^+
+\mathbf{T}^{+-}\mathbf{u}_1^- \ \
\end{eqnarray}
 where $\mathbf{T}^{\pm\pm}$ denote the corresponding
$2\times 2$ - submatrices of the $4\times 4$ - matrix $T$ .

For real $\kappa_1$ the amplitude of the incident wave
propagating  in the region $x_3 <-l/2$ in the positive
$x_3$-direction is $\mathbf{c}_{in}=\mathbf{c}_1^+
=\mathbf{u}_1^+\ex{-\ie\kappa_1 l/2}$. The amplitude of the
reflected wave is $\mathbf{c}_{r}=\mathbf{c}_1^-
=\mathbf{u}_1^-\ex{\ie\kappa_1 l/2}$ and that of the transmitted
wave is given by
$\mathbf{c}_{t}=\mathbf{c}_3^+=\mathbf{v}_2^+\ex{\ie\kappa_3
l/2}$ for real $\kappa_3$.
The amplitudes  $\mathbf{c}_{r}$, $\mathbf{c}_{t}$ are
obtained from (\ref{eqamp}):
\begin{eqnarray}\label{solref1}
\mathbf{c}_{r}=-\ex{\ie\kappa_1 l}
(\mathbf{T}^{--})^{-1}\mathbf{T}^{-+}\mathbf{c}_{in},\ \\
\label{solref2} \mathbf{c}_{t}= \ex{\ie(\kappa_3+\kappa_1)
l/2}[\mathbf{T}^{++} -
\mathbf{T}^{+-}(\mathbf{T}^{--})^{-1}\mathbf{T}^{-+}]
\mathbf{c}_{in}.\
\end{eqnarray}
If $\kappa_3$ is imaginary, then $\mathbf{c}_r$ yields again the
amplitude of the reflected wave (total reflection), whereas
$\mathbf{c}_t$ describes the amplitude of the decaying wave.

If both $\kappa_1$ and $\kappa_3$ are imaginary, then the waves are totally reflected
at both $x_3=\pm l/2$. The waves obey $\mathbf{v}_2^-=\mathbf{u}_1^+=0$.
Then the equations (\ref{eqamp}) can have a nonzero
solution only if $\kappa_3$ is imaginary (since in virtue of
(\ref{Pointing}), $\mathbf{V}_2^{*}G_3\mathbf{V}_2= \mathbf{U}_1^*
G_1\mathbf{U}_1=0$), and $\det \mathbf{T}^{--}=0$ with
\begin{eqnarray}\label{Tmm}
{\bf T}^{--} =\frac{1}{4}( {\bf T}_2^{-+} \ex{-\ie\kappa_2l} {\bf
T}_1^{+-} +{\bf T}_2^{--} \ex{+\ie\kappa_2l} {\bf T}_1^{--})=  \nonumber \\
\nonumber = \frac{1}{4}({\bf T}_2^{--} (\ex{2\ie\kappa_2 l}{\bf
1}-{\bf R}_2 {\bf R}_1 ) \ex{-\ie\kappa_2 l} {\bf T}_1^{--}), \\
\label{Tmm1}{\bf R}_2 = -({\bf T}^{--}_2)^{-1} {\bf T}^{-+}_2, \
{\bf R}_1 = {\bf T}^{+-}_1 ({\bf T}^{--}_1)^{-1}.
\end{eqnarray}
The matrices ${\bf R}_j$ describe the total reflection of the
waves coming from the center to $l_j$, ${\bf v}_1^+ = {\bf
R}_1{\bf v}_1^-$, ${\bf u}_2^-={\bf R}_2{\bf u}_2^+$. These
matrices differ by a similarity transformation from unitary
matrices ${\bf O}_j= {\bf g}_2^{1/2} {\bf R}_j {\bf g}_2^{-1/2}$.
Thus one obtains electromagnetic waves propagating in layer 2 as
soon as one of the two eigenvalues $\ex{\ie\phi}$ of the unitary
matrix ${\bf O}_2{\bf O}_1$ agrees with $\ex{2\ie\kappa_2 l}$.

If $\kappa_j$ is real, then the functions $\mathbf{f}_j^{\,
\pm}(x_3)\ex{\ie \bar{p}\bar{x}}$, describe plane waves
propagating in the medium with constants $\varepsilon_j$, $\mu_j$
in directions of vectors $\vec{p}_j^{\,\pm}=(p_1,p_2,\pm \kappa_j)
$ with velocity $v_j=cp_0/|\vec{p}_j^{\, \pm}|=c/n_j$. For the
angle $\vartheta_j$  between $\vec{p}_j$ and the $x_3$-axis it
holds $\sin\vartheta_j = \mathfrak p/|\vec{p}_j|=\mathfrak
p/(p_0n_j),$ and this equality yields Snell's law
$\sin\vartheta_j/ \sin\vartheta_k=n_k/n_j$. The component
$v_j^{3\pm}$ of the wave front velocity $v_j$ is equal to
$v_j^{3\pm}=\pm v_j\kappa_j/|\vec{p}_{j}^\pm|= \pm c\kappa_j/(p_0
n_j^2)$.

The electric field vector of  the wave propagating in the $j$-th
layer in the positive (negative) direction of the $x_3$ axis is
$\vec{E}_j^{+}=-\ie p_0 \vec{A}^{+}_j$ ($\vec{E}^{-}_j=-\ie p_0
\vec{A}^{-}_j$), and the corres\-ponding energy density is
$\varepsilon_j |\vec{E}^+_j|^2$ ($\varepsilon_j |\vec{E}^-_j|^2)$.
The energy current density propagating in the positive
$x_3$-direction  is $I_j= I_j^{+}-I_j^{-}, \
I_j^{\pm}=v_j^{3+}\varepsilon_j |\vec{E}^\pm_j|^2$. In virtue of
(\ref{amppar}-\ref{amport1}),
\begin{eqnarray}
\nonumber I_j^{\pm} = I_{\rho j}^{\pm}+I_{\tau j}^{\pm},\ I_{\rho
j}^{\pm}=\frac{p_0^3 e_j |\rho^{\pm}_j|^2}{\mathfrak{p}^2},\
I_{\tau j}^{\pm} = \frac{p_0 m_j
|\tau^{\pm}_j|^2}{\mathfrak{p}^2}.
\end{eqnarray}
If we denote $U_3\equiv V_2$, then it holds $I_j=\mathfrak{p}^2
U^*_jG_jU_j/p_0^2$. The energy is conserved in the non-active
medium, therefore the quantity $I_j$ is independent of $x_3$ and
$I_j=I_k$ (in agreement with (\ref{Pointing})). In virtue of
(\ref{Pointing}), the ener\-gy current $I_j$ vanishes in case of
total reflection, since $V_2^*G_3V_2=0$ by imaginary $\kappa_3$ and
${\bf v}_2^-=0$.

If $\kappa_j$ is imaginary, the waves propagate  in the $j$-th layer parallel
to the plane $x_3=0$ in direction of vector $\vec{p}_\|$ similarly as in a
wave-guide. Due to the boundary-conditions given by the matrices ${\bf O}_i$
the relation between $\omega$ and $\vec p_\|$ will be changed.

\section{Conclusion}

The Chern-Simons interaction at $x_3=l_i$ does not change Snell's law.
However, the reflection and transmission coefficients depend on the strengths
$a_i$ of these interactions. They lead to a mixing between the parallel and
perpendicular components of the electromagnetic waves and they change the
relation between frequency and wave-vector for waves between two totally
reflecting media. Consequently such interactions will also modify the strength
of the Casimir effect.
A search for surfaces or layers showing such a behavior
is of high interest.

The presented results may be verified experimentally. In this way,
it is possible  to determine  the constant $a$ describing the
interaction of film with the electromagnetic field in our model.
By finite $a$ the Chern-Simons potential breaks the time and space
parity. It holds also for interaction of photons with
(2+1)-dimensional Dirac field modelling two-dimensional material
\cite{10}$^,$ \cite{11}.

 In this paper we have considered only the case of inactive
medias ( $\Im a_j= \Im \epsilon_j= \Im \mu_j =0$). Using complex
values of the model parameters  and taking also into account the
defect contribution of the (3+1)-dimensional Dirac field \cite{9},
it is possible within Symanzik approach to construct in quantum
electrodynamics a model for wide class of quantum macroscopic
phenomena in systems with two-dimensional space inhomogeneities.
In such a model one can investigate the Hall effect, plasmonics,
nanophotonics, topological insulators, properties of
two-dimensional materials, doping, thin films and sharp
interfaces.

Recently  one places high emphasis on these problems, and many
important results are obtained in studies of them \cite{11}$^,$
\cite{12}. The comprehensive model built within the proposed approach
and based on fundamental physical principles seems to be
suitable for this research field. We expect that it provides an
opportunity to obtain more accurate quantitative results, than
those which have been achieved to date by use of other
theoretical assumptions. Investigations of such models will enable
us to understand more deeply the relationship between different
nano-physical effects.

\begin{acknowledgments}
D.Yu.P. and Yu.M.P. acknowledge Saint Petersburg State University
for a research grant 11.38.660.2013, and are grateful also to ETH
Z\"urich and Ruprecht Karls University Heidelberg for financial
support and kind hospitality.
\end{acknowledgments}

\newpage
\appendix*
\section{Detailed results and comments}
We give an obvious form of matrices used in our calculations. They
are functions of $\check{e}=(e_1,e_2, e_3),\
\check{m}=(m_1,m_2,m_3)$ and can be written as
\begin{eqnarray}\label{ap1}
  {\bf M}(\check{e},\check{m})= \mat{ f(\check{e},\check{m})
& g(\check{e},\check{m})\\
 -p_0^2g(\check{m},\check{e})& f(\check{m},\check{e})}.
\end{eqnarray}
Thus, ${\bf M}$ is completely defined by its elements $\{{\bf
M}\}_{11}=f(\check{e},\check{m})$ and $\{{\bf
M}\}_{12}=g(\check{e},\check{m})$.

The matrices ${\bf T}_j^{\pm\pm}$ and their inverses read
\begin{eqnarray*}
 \{{\bf T}_j^{\alpha\beta}\}_{11} =
1+\alpha\beta\frac{e_j}{e_{j+1}},\ \{{\bf
T}_j^{\alpha\beta}\}_{12} =
-\alpha\frac{2a_j}{e_{j+1}p_0},\\
\{({\bf T}_j^{\alpha\beta})^{-1}\}_{11}=
\frac{1+\alpha\beta\frac{m_j}{m_{j+1}}}{\det({\bf
T}_j^{\alpha\beta})}, \{({\bf T}_j^{\alpha\beta})^{-1}\}_{12} =
\frac{\alpha\frac{2a_j}{e_{j+1}p_0}}{\det({\bf
T}_j^{\alpha\beta})},\\
\det({\bf T}_j^{\alpha\beta}) = \frac{4a_j^2+(e_{j+1}+\alpha\beta
e_j)(m_{j+1}+\alpha\beta m_j)}{e_{j+1}m_{j+1}}.
\end{eqnarray*}
The matrices $\mathbf{T}^{\pm\pm}$ obey
\begin{eqnarray*}
 \mathbf{T}^{\pm\pm}= \cos(\kappa_2 l)
 \mathbf{Z}^{\pm\pm}_1+\ie \sin(\kappa_2 l)
 \mathbf{Z}^{\pm\pm}_2
, \\
\{\mathbf{Z}^{\alpha\beta}_1\}_{11}=\frac{\alpha \beta
e_1+e_3}{2e_3}, \{\mathbf{Z}^{\alpha\beta}_1\}_{12}=-\frac{\alpha (a_1+a_2)}{e_3p_0}, \\
\{\mathbf{Z}^{\alpha\beta}_2\}_{11}= \frac{4 \alpha  a_1 a_2
e_2-(\alpha e_2^2+\beta
 e_1 e_3) m_2}{2e_2m_2e_3},\\
 \{\mathbf{Z}^{\alpha\beta}_2\}_{12}
= \frac{\alpha \beta a_2 e_2 m_1+a_1 e_3 m_2}{e_2m_2e_3p_0},\
\alpha,\beta=\pm 1.
\end{eqnarray*}

The relations (\ref{solref1},\ref{solref2}) for the amplitudes
$\mathbf{c}_t,\ \mathbf{c}_r$ can be written as
$\mathbf{c}_r=-\ex{\ie\kappa_1 l} \mathbf{T}_r\mathbf{c}_{in}$, $
  \mathbf{c}_t=\ex{\ie(\kappa_3+\kappa_1)l/2}
  \mathbf{T}_t\mathbf{c}_{in}$ with
\begin{eqnarray*}
\mathbf{T}_r =(\mathbf{T}^{--})^{-1}\mathbf{T}^{-+},
  \mathbf{T}_t=
\mathbf{T}^{++} -
\mathbf{T}^{+-}(\mathbf{T}^{--})^{-1}\mathbf{T}^{-+}.
\end{eqnarray*}
Using the notations
\begin{eqnarray*}
\varphi(a,b)= a\cos(\kappa_2 l)+\ie\, b \sin(\kappa_2 l),
\\
\psi(a,b,c)=b(a+c)\cos(\kappa_2l)+\ie\,(ac+b^2)\sin(\kappa_2l),
\\
  e_i^\alpha=\varphi(\alpha e_2,e_i),\ m_i^\beta=\varphi(\beta m_2,m_i),\
  \varphi_{i}^{\alpha\beta}=e_i^\alpha m_i^\beta,\\
  e^{\alpha}=\psi(e_1,\alpha e_2,\alpha e_3),m^{\beta}=\psi(m_1,\beta
  m_2,\beta,
  m_3),\\ \psi^{\alpha\beta}=e^\alpha m^\beta,\ \alpha,\beta=\pm
  1,
\end{eqnarray*}
one can present the matrices $\mathbf{T}_t$, $\mathbf{T}_r$ in the
following form
\begin{eqnarray*}
\{\mathbf{T}_t\}_{11}= \frac{2e_1(e_2 m^+  -4\ie\,
a_1 a_2 m_2 \sin(\kappa_2 l))}{z} ,\\
\{\mathbf{T}_t\}_{12}=-\frac{4m_1(a_2m_2 e_1^+ + a_1 e_2 m_3^+)}{p_0z},\\
\{\mathbf{T}_r\}_{11}= \frac{1}{z}(8 a_1a_2 e_2 m_2
+\psi^{-+}+\\
+4(
 a_1^2\varphi_3^{++}-a_2^2 \varphi_{1}^{-+} - 4a_1^2 a_2^2 \sin^2(\kappa_2 l))
 ),\\
 \{\mathbf{T}_r\}_{12}=\frac{4m_1( a_2e_2m_2+a_1 (\varphi_3^{++}-4
a_2^2 \sin^2(\kappa_2 l))}{ p_0z} ,
\end{eqnarray*}
where
\begin{eqnarray*}
 z= 4e_2m_2e_3m_3\det\mathbf{T}^{--} =\psi^{++}+8 a_1a_2e_2m_2 +\\
 + 4(a_2^2 \varphi_1^{++}+ a_1^2
 \varphi_3^{++}-4a_1^2a_2^2\sin^2(\kappa_2 l)).
\end{eqnarray*}

The reflection matrices ${\bf R}_i$ defined by (\ref{Tmm1}) are
\begin{eqnarray*}
{\bf R}_1 =
{\bf T}^{+-}_1 ({\bf T}^{--}_1)^{-1},\ {\bf R}_2 = -({\bf T}^{--}_2)^{-1} {\bf T}^{-+}_2, \\
 \{{\bf R}_i\}_{11}= -\frac{r^{-+}_i}{r^{++}_i},\ \{{\bf
R}_i\}_{12}= -  \frac{4 a_im_2}{r^{++}_i}
,\\
r^{\alpha\beta}_1=4a^2_1+(e_1+\alpha e_2)(m_1+\beta m_2),\\
r^{\alpha\beta}_2=4a^2_2+(e_3+\alpha e_2)(m_3+\beta m_2),\
\alpha,\beta =\pm 1.
\end{eqnarray*}
Multiplication and the inverse of matrices of the form
(\ref{ap1}) yield matrices of the same type. Because
${\bf g}_2$ does not belong to this class of matrices, this is also
the case for the matrices $ {\bf O}_j = {\bf g}_2^{1/2} {\bf R}_j
{\bf g}_2^{-1/2}$:
\begin{equation}\label{ap2}
{\bf O}_j= -\frac{1}{r_j^{++}}\mat{r_j^{-+}
& 4 a_j \sqrt{e_2 m_2} \\
-4 a_j \sqrt{e_2 m_2} & r_j^{+-} }.
\end{equation}

The $r^{\pm\pm}$ obey
\begin{eqnarray}\label{ap3}
r^{+-}_jr^{-+}_j +16 a^2_j e_2m_2=r^{++}_jr^{--}_j.
\end{eqnarray}
If $a_1,a_2,e_2,m_2$ are real, and $e_1,e_3,m_1,m_3$ are
imaginary, then $(r^{-+})^*= r^{+-}$, $(r^{++})^*= r^{--}$, and it
follows from (\ref{ap2},\ref{ap3}) that the matrices ${\bf O}_1,
{\bf O}_2$ and
\begin{eqnarray}
{\bf O}={\bf O}_2{\bf O}_1=\frac{1}{R}\mat{P & Q \\
-Q^* & P^* } \nonumber
\end{eqnarray}
with
\begin{eqnarray*}
R=r_1^{++}r_2^{++},\ P=r_1^{-+}r_2^{-+}-16 a_1a_2e_2m_2,\\ Q=
4\sqrt{e_2m_2}(a_1r_2^{-+}+ a_2r_1^{+-}),\ PP^*+QQ^*=RR^*
\end{eqnarray*}
 are unitary.

The eigenvalues $\lambda_{1,2}$ of the matrix ${\bf O}$ read
\begin{eqnarray*}
\lambda_{1,2}=\frac{-P-P^*\pm
\sqrt{(P-P^*)^2-4QQ^*}}{2R}=e^{i(\zeta+\eta_{1,2})},\\
\tan (\zeta)=-\frac{\Im R}{\Re R},\ \tan
(\eta_{1,2})=\mp\frac{\sqrt{(\Im P)^2+|Q|^2}}{\Re P}.
\end{eqnarray*}
They coincide for $\Im P=0, Q=0$. In this case $\eta_{1,2}=0$,
\begin{eqnarray*}
r_2^{-+}=-\frac{a_2}{a_1}r_1^{+-}, \
P=-\frac{a_2}{a_1}r_1^{++}r_1^{--}=P^*.
\end{eqnarray*}

The boundary conditions (\ref{discD}-\ref{discH}) can be proved
directly from (\ref{eqa3},\ref{VPot}). Using the relations
$\vec{D}=\varepsilon\vec{E}$, $\vec{B}=\mu\vec{H}$,
$\vec{E}=-\partial_0\vec{A}$, $\vec{B}=\vec{\partial}\times
\vec{A}$,
 $\mathfrak p^2+\kappa^2=p_0^2\varepsilon\mu$ and
notations $\varepsilon\kappa=e,\ \kappa/ \mu=m $, we obtain $D_3
=-p_0e \tilde{\rho}$,
\begin{eqnarray}
\nonumber H_1 =  \frac{p_1 m \tilde{\tau}-p_2e \tilde{\rho}
p_0^2}{\mathfrak p^2} , \ H_2 = \frac{p_1e \tilde{\rho}
p_0^2+p_2m\tilde{\tau} }{\mathfrak p^2} .
\end{eqnarray}
It follows from $J^{\nu} = \epsilon^{3 \nu \sigma \rho}F_{\sigma
\rho}$ that $J^0 = 2\tau$,
\begin{eqnarray}\nonumber
\ J^1 = 2\frac{p_0(p_1\tau-p_2\rho)}{\mathfrak p^2} ,\ J^2 = 2
\frac{p_0(p_1\rho+p_2\tau)}{\mathfrak p^2}.
\end{eqnarray}
Thus in virtue of (\ref{cond1a}), the equalities
(\ref{discD}-\ref{discH}) are fulfilled.
\end{document}